\documentclass[letters,usenatbib]{mnras}

\usepackage{newtxtext,newtxmath}

\usepackage[T1]{fontenc}

\DeclareRobustCommand{\VAN}[3]{#2}
\let\VANthebibliography\thebibliography
\def\thebibliography{\DeclareRobustCommand{\VAN}[3]{##3}\VANthebibliography}

\usepackage{amsmath}	

\usepackage{academicons}
\usepackage{hyperref}

\usepackage{svg}

\newcommand{\farc}{$^{\prime\prime}$}

\title[Infrared quasars buried in compact starbursts]{Obscuration beyond the nucleus: infrared quasars can be buried in extreme compact starbursts}

\author[C. Andonie et al.]{Carolina Andonie
,$^{1}$\thanks{E-mail: carolina.p.andonie@durham.ac.uk}
David M. Alexander,$^{1}$
Claire Greenwell,$^{1}$
Annagrazia Puglisi,$^{2,1}$\thanks{Anniversary Fellow}
Brivael Laloux,$^{3,1}$
\newauthor
Alba V. Alonso-Tetilla,$^{2}$
Gabriela Calistro Rivera, $^{4}$
Chris Harrison,$^{5}$
Ryan C. Hickox,$^{6}$
Melanie Kaasinen,$^{4}$
\newauthor
Andrea Lapi,$^{7}$
Iv\'an E. L\'opez,$^{8,9}$
Grayson Petter,$^{6}$
Cristina Ramos Almeida,$^{10,11}$
David J. Rosario,$^{5}$
\newauthor
Francesco Shankar,$^{2}$
and Carolin Villforth$^{12}$\\
$^{1}$ Centre for Extragalactic Astronomy, Department of Physics, Durham University, Durham, DH1 3LE, UK\\
$^{2}$ School of Physics and Astronomy, University of Southampton, Highfield SO17 1BJ, UK\\
$^{3}$Institute for Astronomy \& Astrophysics, National Observatory of Athens, V. Paulou \& I. Metaxa 11532, Athens, Greece\\
$^{4}$European Southern Observatory (ESO), Karl-Schwarzschild-Straße 2, 85748 Garching bei München, Germany\\
$^{5}$School of Mathematics, Statistics and Physics, Newcastle University, Newcastle upon Tyne, NE1 7RU, UK\\
$^{6}$ Department of Physics and Astronomy, Dartmouth College, 6127 Wilder Laboratory, Hanover, NH 03755, USA\\
$^{7}$ SISSA, Via Bonomea 265, I-34136 Trieste, Italy\\
$^{8}$Dipartimento di Fisica e Astronomia  "Augusto Righi", Universit\`a di Bologna, Via Gobetti 93/2, 40129 Bologna, Italy\\
$^{9}$ INAF – Osservatorio di Astrofisica e Scienza dello Spazio di Bologna, Via Gobetti 93/3, 40129 Bologna, Italy\\
$^{10}$ Instituto de Astrof\'isica de Canarias, Calle V\'ia L\'actea, s/n, 38205 La Laguna, Tenerife, Spain \\
$^{11}$ Departamento de Astrof\'isica, Universidad de La Laguna, 38206 La Laguna, Tenerife, Spain \\
$^{12}$ Department of Physics, University of Bath, Claverton Down, Bath BA2 7AY, UK \\
}

\date{Accepted XXX. Received YYY; in original form ZZZ}

\pubyear{2023}
\defcitealias{2022Andonie}{A22}
\begin{document}
\label{firstpage}
\pagerange{\pageref{firstpage}--\pageref{lastpage}}
\maketitle

\begin{abstract}
In the standard quasar model, the accretion disk obscuration is due to the canonical dusty torus. Here, we argue that a substantial part of the quasar obscuration can come from the interstellar medium (ISM) when the quasars are embedded in compact starbursts. We use an obscuration-unbiased sample of 578 infrared (IR) quasars at $z\approx 1-3$ and archival ALMA submillimeter host galaxy sizes to investigate the ISM contribution to the quasar obscuration. We calculate SFR and ISM column densities for the IR quasars and a control sample of submillimeter galaxies (SMGs) not hosting quasar activity and show that: (1) the quasar obscured fraction is constant up to $\rm SFR\approx 300 \: M_{\odot} \: yr^{-1}$, and then increases towards higher SFR, suggesting that the ISM obscuration plays a significant role in starburst host galaxies, and (2) at $\rm SFR\gtrsim 300 \: M_{\odot} \: yr^{-1}$, the SMGs and IR quasars have similarly compact submillimeter sizes ($R_{\rm e}\approx 0.5-3\rm \: kpc$) and, consequently, the ISM can heavily obscure the quasar, even reaching Compton-thick ($N_{\rm H}>10^{24} \rm \: cm^{-2}$) levels in extreme cases. Based on our results, we infer that $\approx 10-30\%$ of the IR quasars with $\rm SFR\gtrsim 300 \: M_{\odot} \: yr^{-1}$ are obscured solely by the ISM.

\end{abstract}

\begin{keywords}
quasars: general -- galaxies: starburst -- submillimetre: galaxies
\end{keywords}



\section{Introduction}

Quasars are the most luminous subset of the Active Galactic Nucleus (AGN) population and represent the most rapid phase of supermassive black hole (SMBH) growth. Strong observational evidence supports a connection between SMBH accretion and the evolution of their host galaxies  \citep{2012Alexander,2013Kormendy,2014Madau}, which is expected to occur in an early obscured phase \citep{1988Sanders,2008aHopkins,2014Lapi, 2018HickoxyAlexander}, with large reservoirs of gas and dust feeding the SMBH and the galaxy. In the last few years, there has been growing evidence favouring this synchronized evolution due to the discovery of fundamental differences in the host galaxy and nuclear properties between obscured ($N_{\rm H}>10^{22}\rm \: cm^{-2}$) and unobscured quasars ($N_{\rm H}<10^{22}\rm \: cm^{-2}$), which cannot be easily explained by the standard AGN orientation model \citep{1993Antonucci, 1995Urry}; i.e., the observed differences cannot be only due solely to the orientation of a geometrically and optically thick dusty torus \citep{2017RamosAlmeida}. A number of recent studies have found systematic differences between obscured and unobscured quasars in the clustering properties \citep{2011Hickox,2014DiPompeo,2014Donoso,2014Villarroel,2018Powell,2023Petter}, star-formation (SF) properties \citep{2015Chen,2022Andonie}, and galaxy companion environments \citep{2014Satyapal,2017Weston,2023Dougherty}. Combined, these results suggest that obscured quasars represent a distinct early growth phase closely connected to a period of enhanced SF.

Recently, \citet{2022Andonie} (hereafter \citetalias{2022Andonie}) found that obscured quasars have star-formation rates (SFR) $\approx 3$ times higher than unobscured quasars, a result inconsistent with the standard AGN orientation model. Most striking is the finding that an excess of obscured quasars reside in starburst galaxies with $\rm SFR>300 \: M_{\odot} \: yr^{-1}$, providing evidence that the host galaxy may contribute to the total obscuration of the quasar. A number of previous studies have investigated the host galaxy contribution to the total AGN obscuration \citep[e.g.,][]{2007Gilli, 2017Buchner, 2019Circosta,Gilli22,2023Silverman}, but our current study is the first to directly constrain the host galaxy obscuration as a function of the SFR. To achieve this we use observations from the Atacama Large Millimeter/submillimeter Array (ALMA) of a sub sample of the \citetalias{2022Andonie} IR quasars. We use these ALMA data to improve the SFR constraints and explore the host galaxy submillimeter (submm) sizes to assess the contribution of the interstellar medium (ISM) to the total quasar obscuration. We also analyse a control sample of submm galaxies (SMGs), which do not host an AGN. 

In this work, we adopt a concordance cosmology \citep{2013Hinshaw}, and a Chabrier initial mass function \citep[][]{2003Chabrier}.

\section{Observations and analysis}

\subsection{SED fitting, photometric catalogues, and the sample} \label{subsec:sedfitting}

Our parent sample is taken from the quasar catalogue published by \citetalias{2022Andonie} from the Cosmic Evolution Survey (COSMOS, \citealp[][]{2007Scoville}). \citetalias{2022Andonie} identified a complete sample of 578 infrared (IR) quasars ($L_{\rm AGN,8-1000\: \mu m}>10^{45} \rm \: erg \: s^{-1}$) at $z<3$, with minimal obscuration bias, using detailed UV-to-far IR spectral energy distribution (SED) fitting. We find that 24 IR quasars (hereafter the submm-IR quasars) have archival ALMA host galaxy submm size measurements (see Section\,\ref{subsec:alma} for details). In this study we re-fit the rest-frame $0.3-1000 \rm \: \mu m$ SEDs of the submm-IR quasars but now including the ALMA Band 6/7 photometry from the latest version of the A3COSMOS photometric catalogue (version 20200310, \citealp{2019LiuD}). Our SED fitting is performed using the multicomponent Bayesian SED fitting code \textsc{fortesfit} \footnote{https://github.com/vikalibrate/FortesFit} \citep[][]{2019Rosario}, and the same models and SED fitting approach of \citetalias{2022Andonie}. The SED modelling includes the unabsorbed stellar emission, the UV emission from the accretion disk, IR emission from the AGN, and the dust obscured star-formation (for more details, see Section~2 and~3 of \citetalias{2022Andonie}). As a comparison, we also fit a control sample of all the 85 SMGs with reliable host galaxy sizes in COSMOS reported by \citet{2017Fujimoto}, using the same SED fitting approach as for the IR quasars. We exclude from the SMG sample all the sources either identified as AGN by our SED fitting, detected in the {\it Chandra} COSMOS Legacy survey (\citealp{2016Civano}, see Section\,\ref{subsec:Xray}), or part of the AGN optical spectroscopic catalogue of \citet{2013Rosario}, ending up with 78 sources. For details on the AGN identification process and SFR upper-limits estimation, we refer the readers to Section\,3.3 of \citetalias{2022Andonie}.  The SED best-fitting parameters are reported in Table\,1 of the online Supplementary Material.

\begin{figure}
\centering
\includegraphics[trim={2.6cm 3.cm 3.5cm 2.8cm},clip,scale=0.2]{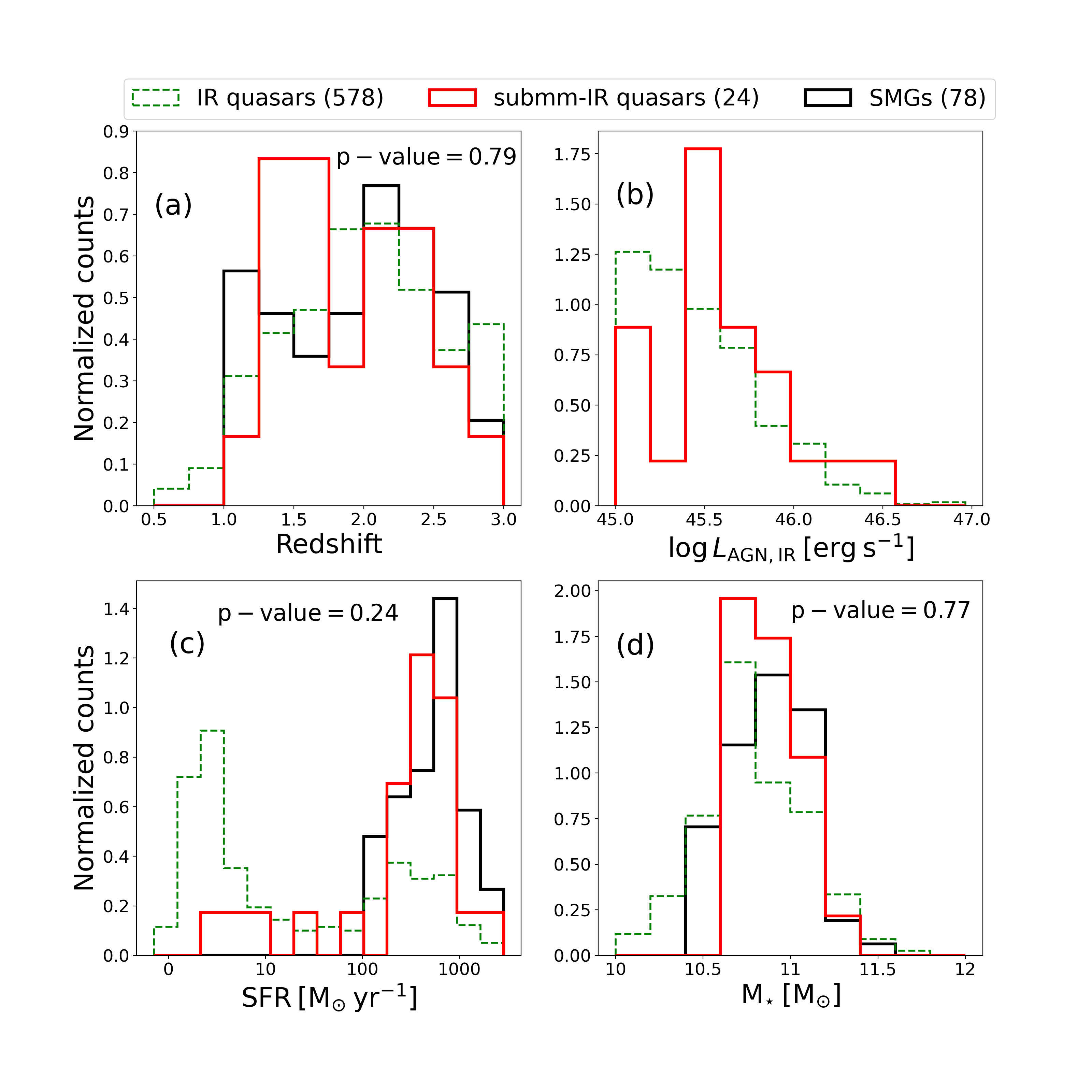}

\caption{Normalized distributions for the full IR quasar sample (578 sources) of \citetalias{2022Andonie} (green dotted), the 24 IR quasars with ALMA sizes (red solid), and the control sample of 78 SMGs (black solid). In 3 panels, the K-S $\rm p-value$ between the IR quasars with sizes and the SMGs is quoted. (a): redshifts; (b): $8-1000\: \rm \mu m$ AGN luminosity (c): SFRs; (d) stellar masses. } 
\label{fig:prop}
\end{figure}

Figure\,\ref{fig:prop} depicts the host galaxy properties of the IR quasar sample of \citetalias{2022Andonie}, the submm-IR quasars, and our control sample of SMGs, and also reports the Kolmogorov Smirnov (KS) test p-value between the submm-IR quasars and SMGs. We plot the 50th percentile from the posterior distribution for the SED best-fitting parameters. Panels (a) and (d) show that the three samples have consistent redshifts and stellar masses. Similarly, panel (c) shows that the SFRs of the IR quasars span a wide range of values, and that the submm-IR quasars are a good representation of the IR quasars with the largest SFRs; the distributions of the redshifts, SFRs, $\rm M_{\star}$, and $L_{\rm AGN, IR} $ of the submm-IR quasars and the IR quasars with $\rm SFR >100 \:M_{\odot} \: yr^{-1}$  always have p-values of $>0.3$. However, we note that the majority of the IR quasars have low SFRs ($<10\rm \: M_{\odot}\:yr^{-1}$), and we can only provide an SFR upper limit in these cases; hence, those sources are excluded from the KS test analysis. We find that the distributions of the SFRs and stellar masses of the submm-IR quasars (mean and standard deviation values of $\rm SFR_{submm-IR\: quasars}= 610\pm 400 \: M_{\odot} \: yr^{-1}$ and $M_{\rm \star, submm-IR \: quasars}=(8\pm 4)\cdot 10^{10} \rm \: M_{\odot}$) and SMGs (mean and standard deviation values of $\rm SFR_{SMGs}= 700\pm 550 \: M_{\odot} \: yr^{-1}$ and $M_{\rm \star, SMGs}=(8.6\pm 4.2)\cdot 10^{10} \rm \: M_{\odot}$) are also similar (in both cases the KS p-value $>0.2$) and place both populations on the starburst region of the main-sequence of star-forming galaxies, following \citet{2012Whitaker} and \citet{2019Aird}.

\subsection{ALMA data} \label{subsec:alma}

We use submm (i.e., dust continuum) measurements from the Band 6 or 7 ($\sim 1\rm \: mm$, whichever is available) ALMA data of \citet{2017Fujimoto}, \citet{2020Chang}, and \citet{2021Lamperti}, in addition to measurements from a combination of CO(5-4), CO(4-3), CO(2-1), and $1.1-1.3 \rm \: mm$ continuum observations from \citet{2019Puglisi,2021Puglisi}\footnote{There is no bias in the sizes from \citet{2019Puglisi} as they showed that the CO(5-4), CO(4-3) are consistent with $\sim 1\rm \:mm$ sizes for their sample.}. Overall, there are 24 IR quasars with reliable host galaxy sizes from ALMA observations with $\rm S/N>5$, and all of them used a circularized Gaussian profile, or a consistent geometry\footnote{\citet{2021Lamperti} fitted different models to measure sizes, and for the three overlapping sources from our sample, the best-fitting model sizes are consistent within $1\sigma$ with those derived using a circularised Gaussian profile.}, to fit the data. Here, we define the submm effective radius as $R_{e}=\rm FWHM/2$. The size measurements and references for each source in our sample are reported in Table\,1 of the Supplementary Material.

The angular resolution of the observations used to calculate the sizes covers $\theta_{\rm ALMA}\approx$ 0.\farc{}18-0.\farc{}9 ($\approx 1.5-7 \rm \: kpc$ at $z=1-3$), including 6 out of 24 objects with $\theta_{\rm ALMA}<0.$\farc{}3. We note that we do not expect any of these observations to over-resolve the galaxy, given that most of the dust continuum emission is typically concentrated into a central compact component of $\sim$0.\farc$15$ \citep{2019Gullberg}.

The AGN-heated dust could affect the host galaxy size measurements if its contribution to the ALMA Band 6/7 flux is significant. However, on the basis of our SED decomposition (see Section\,\ref{subsec:sedfitting} for details), the AGN contribution to the total rest-frame $850 \rm \: \mu m$ luminosity is $>10\%$ for only three sources and $<5\%$ for 19 sources. For the three sources with the slightly largest quasar contribution, the host galaxy sizes are $>1.3\rm \: kpc$, which is larger than the average size of the submm-IR quasars (see Section\,\ref{subsec:ISMNH} for details). Hence, we conclude that the quasar emission has a minimal impact on the host galaxy submm size measurements.

\subsection{X-ray spectral constraints} \label{subsec:Xray}

In the \citetalias{2022Andonie} sample, 63\% (365/578) of the IR quasars are X-ray detected in the {\it Chandra} COSMOS Legacy survey \citep{2016Civano}, while the other 37\% remain undetected. In this work, we adopted the X-ray spectral fitting constraints from \citet{2023Laloux}, unlike \citetalias{2022Andonie}, who used the measurements from \citet{2016MarchesiXray} and \citet{2018Lanzuisi}. For the overlapping sources, both catalogues are in broad agreement with the absorbing column densities for sources with $N_{\rm H}<10^{23} \rm \: cm^{-2}$, and the classification between obscured/unobscured sources does not change. However, \citet{2023Laloux} used a Bayesian approach that robustly constrains the X-ray spectral physical parameters for sources with  $<30$ counts, which lie below the lower count threshold adopted in \citet{2016MarchesiXray}. Analyzing such faint sources is fundamental in this study, justifying our choice of X-ray catalogue. 

Briefly, in \citet{2023Laloux}, each source is fitted using the BXA package \citep{Buchner_2014} with the UXCLUMPY X-ray model \citep{Buchner_2019}. The output of the fit is a parameter posterior distribution that fully encapsulates the measurement uncertainties. We calculate the probability that an individual source is obscured by taking the fraction of the line of sight (l.o.s) column density posterior distribution above $N_{\rm H,X \: (los)}=10^{22} \rm \: cm^{-2}$, $f_{N_{\rm H}>22}$. In the case of the X-ray undetected sources, we set $f_{N_{\rm H}>22}=1$ following the results of \citetalias{2022Andonie}, who used a combination of X-ray stacking and $L_{\rm 2-10\rm keV}$ -- $L_{\rm 6\: \mu m}$ analyses to demonstrate that the majority ($>$~85\%) of the X-ray undetected sources are heavily obscured by $N_{\rm H,X \: (los)}>10^{23} \rm \: cm^{-2}$, and many are likely Compton-thick ($N_{\rm H,X \: (los)}>10^{24} \rm \: cm^{-2}$, CT); see Section\,4.3, and Figures\,6 and\,7 of \citetalias{2022Andonie} for details. We note that we do not expect the stacked X-ray signal to be contaminated by SF processes\footnote{Adopting \citet{2003Ranalli}, the X-ray luminosity due to $\rm SFR>300 \: M_{\odot} \:yr^{-1}$ ranges $L_{\rm 2-10 \: keV, SF}=(2-9)\cdot 10^{42} \: \rm erg \: s^{-1}$, which is $<5\%$ of the AGN X-ray luminosity calculated from the $L_{\rm 6\: \mu m}$ following \citet{2015Stern}.}, but to further confirm that the X-ray undetected IR quasars with large SFRs are obscured, we take the same approach of \citetalias{2022Andonie} and stack all the X-ray undetected quasars with $\rm SFR>300 \: M_{\odot} \:yr^{-1}$ using the python code \textsc{StackFast}\footnote{The \textsc{StackFast} code is available here \url{https://www.tonima-ananna.com/research}}  (see section 4.3 of that paper). As in \citetalias{2022Andonie}, we detect a strong stacked X-ray signal. The count rates in the soft ($0.5-2\rm \: keV$) and hard ($2-7 \rm \: keV$) bands are $\rm S=(9.95\pm 1.84)\times 10^{-6} \: counts\: s^{-1}$ and $\rm H=(7.73\pm 1.82)\times 10^{-6} \: counts\: s^{-1}$, which is equivalent to a hardness ratio of $\rm HR = \frac{H-S}{H+S} = 0.13 \pm 0.15$. At the mean redshift of the sample ($z=2.2\pm 0.5$), the HR indicates that, on average, these IR quasars are obscured by $N_{\rm H}>10^{23} \rm \: cm^{-2}$ (see Figure\,6 in \citetalias{2022Andonie}) and, consequently, that the majority of the X-ray undetected IR quasars with $\rm SFR>300 \: M_{\odot} \:yr^{-1}$ are X-ray undetected because the X-ray emission was absorbed by heavily-obscured column densities\footnote{The X-ray undetected IR quasars with $\rm SFR<300 \: M_{\odot} \:yr^{-1}$ have a $\rm HR=-0.1\pm0.2$ and $z\approx2$, which also implies $N_{\rm H}>10^{23} \rm \: cm^{-2}$.}.

For the nine submm-IR quasars that are undetected in the X-ray band, we calculated $N_{\rm H,X \: (los)}$ lower limits. For this, we adopt the \cite{2015Stern} relationship to convert the rest-frame 6$\mu m$ AGN luminosity calculated from the SED fits into the intrinsic $2-10 \: \rm keV$ luminosity. We then utilize the {\it Chandra} sensitivity maps generated in \cite{2023Laloux} and use UXCLUMPY to compute the minimum $N_{\rm H,X \: (los)}$ that each source needs to be undetected in the X-ray band at its redshift, which corresponds to the value for what the detection probability drops below 90\%. The $N_{\rm H,X \: (los)}$ ranges $N_{\rm H,X \: (los)}=(0.3-5)\cdot 10^{23} \rm \: cm^{-2}$ and are reported in Table\,1 of the Supplementary Material.

\section{Results }

\subsection{Connection between the obscured fraction and the star-formation rate} \label{subsec:ObsFracSFR}

The SFR enhancement in obscured quasars found in \citetalias{2022Andonie} suggests that dust-obscured SF from the host galaxy may cause at least part of the obscuration towards the quasars. In this section, we demonstrate the connection between obscuration and SFR. 

\begin{figure}
\centering
\includegraphics[trim={0 1.4cm 0 1cm},clip, scale=0.4]{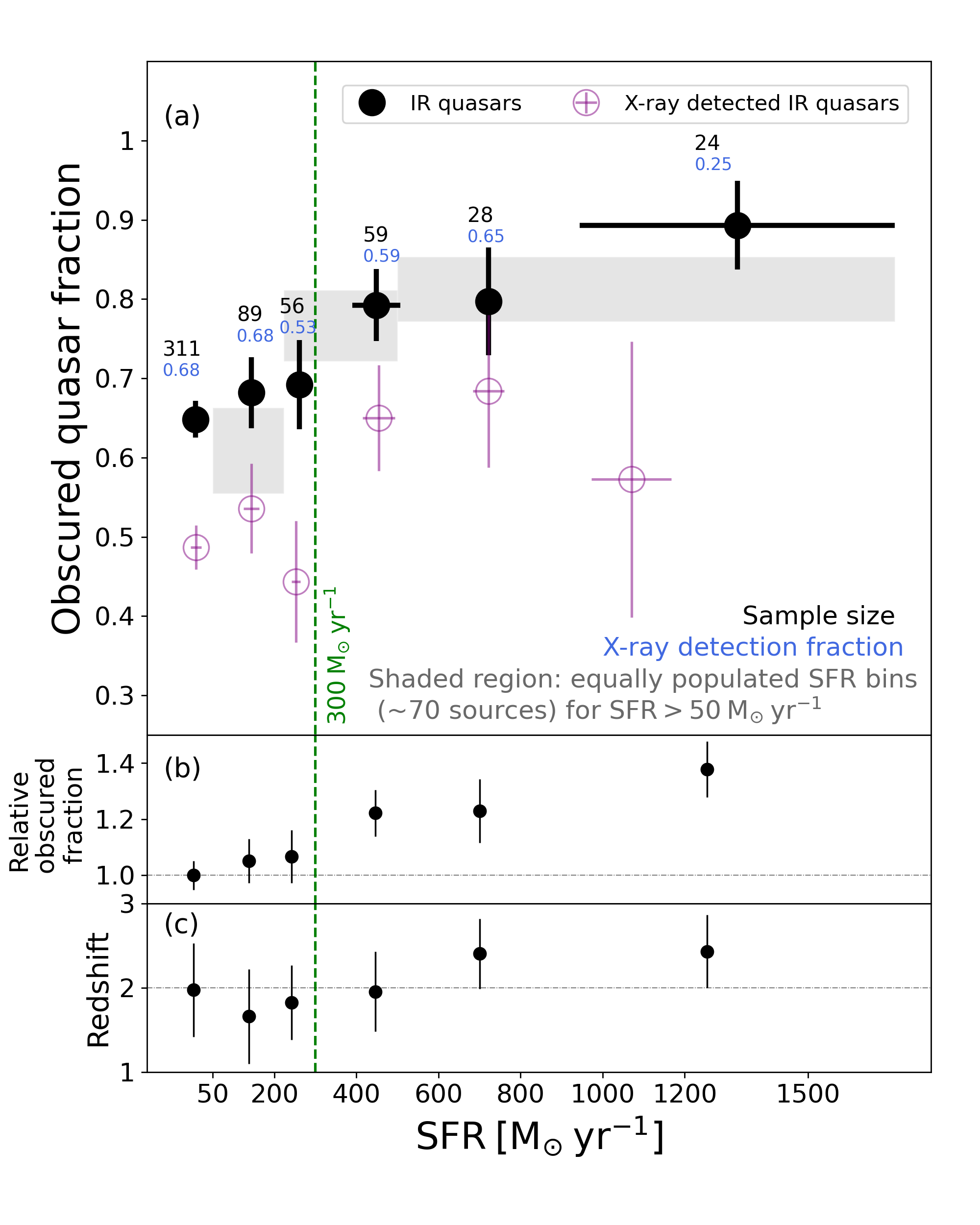}

\caption{{\it Panel (a)}: Obscured quasar fraction versus SFR for the full IR quasar sample of \citetalias{2022Andonie}. Each obscured fraction value corresponds to the mean and standard error of $f_{N_{\rm H}>22}$ at each SFR bin, defined as $\rm SFR=[<100, 100-200, 200-300, 300-600, 600-900,>900]\rm \: M_{\odot} \: yr^{-1}$. Black points correspond to the IR quasars and magenta points are the X-ray detected quasars only. The grey area represents 1$\sigma$ limits for the obscured fraction of three equal-size SFR bins for sources with $\rm SFR>50 \: M_{\odot} \: yr^{-1}$. The black and blue text on top of each point corresponds to the sample size and X-ray detection fraction for each SFR bin, respectively. The x-axes are the mean and standard deviation of the SFRs in each bin. {\it Panel (b)}: Fractional increase of the obscured fraction with respect to the obscured fraction at the lowest SFR bin. {\it Panel (c)}: mean and standard deviation of the redshift distribution at each SFR bin.}

\label{fig:ObsFracSFRs}
\end{figure}

Panel (a) of Figure\,\ref{fig:ObsFracSFRs} shows the obscured quasar fraction as a function of the SFR for the entire IR quasar sample of \citetalias{2022Andonie}. The obscured fraction values and uncertainties correspond to the mean and standard error of the $f_{N_{\rm H}>22}$ distribution at each SFR bin (see Section\,\ref{subsec:Xray} for details). Similarly, the x-axis shows the mean and standard deviation of the SFR distribution at each bin, calculated using \textsc{PosteriorStacker}\footnote{\url{https://github.com/JohannesBuchner/PosteriorStacker}. This code takes into account the upper limits by modelling samples of the posterior distributions using a Gaussian model.}. The total obscured fraction of the sample is $68\pm 2 \%$. However, we find that the obscured fraction has a clear dependence with SFR, being broadly constant ($\approx 65-70\%$) for $\rm SFR< 300 \: M_{\odot} \:yr^{-1}$, increasing to $\approx 80\%$ for $\rm SFR = 300-900 \: M_{\odot} \:yr^{-1}$, and reaching $\approx 90\%$ for $\rm SFR>900 \: M_{\odot} \:yr^{-1}$. Panel (b) clearly illustrates this increase by showing that the obscured fraction at $\rm SFR>300 \: M_{\odot} \:yr^{-1}$ is between $1.2\pm 0.1$ and $1.4\pm 0.1$ times larger than the obscured fraction at the lowest SFR bin.  We note that the number of sources contributing to each SFR bin is not constant: the majority (311/578) reside in the lowest SFR bin, while the two largest SFR bins are comprised of $<30$ sources. However, we still observe the same trend when taking three equally-populated bins ($\approx$ 70 sources) for the quasars with $\rm SFR>50 \: M_{\odot} \:yr^{-1}$ (see the grey area on the graph). The observed increase in the obscured fraction at $\rm SFR> 300 \: M_{\odot} \:yr^{-1}$ suggests that at those SF levels, the host galaxy starts to significantly contribute to the total obscuration of the system.

Panel (c) of Figure\,\ref{fig:ObsFracSFRs} plots the mean and standard deviation of the redshift distributions at each SFR bin, with redshifts ranging $\sim 1.7-2.4$. A general trend of a modest redshift increase with SFR is seen. However, we find that the change in the obscured fraction with SFR is unlikely to be explained by evolution in the intrinsic obscured AGN fraction since this is expected to increase by only up to $\sim 1.1\times$ between $\rm z=[1.5,2.5]$ \citep[e.g.,][]{Gilli22,2023Laloux}. We also note that the stellar mass is broadly constant (median values ranging $M_{\rm \star} \approx (5-7) \cdot 10^{10} \rm \: M_{\odot}$) across the different SFR bins. Hence, the increment of the obscured fraction with SFR is unlikely to be due to a redshift or stellar mass evolution.

We note from Figure\,\ref{fig:ObsFracSFRs} that the X-ray detection fraction does not significantly vary with the SFR, except at the highest SFR bin, dropping from $\approx 0.55-0.70$ to $\approx 0.25$; therefore, the obscured fraction at the largest SFR bin is dominated by X-ray undetected sources. 
In Section\,\ref{subsec:Xray}, we provide strong evidence that the majority of the X-ray undetected sources are obscured by $N_{\rm H}>10^{23} \rm \: cm^{-2}$; hence, we set $f_{N_{\rm H}>22, \rm X\:und}=1$ for the X-ray undetected sources. However, we still observe the same trend between the obscured fraction and the SFR when we conservatively consider that only 75\% of the X-ray undetected quasars are obscured by $N_{\rm H}>10^{22} \rm \: cm^{-2}$ (i.e., $f_{N_{\rm H}>22, \rm X\:und}=0.75$). We finally note that the same trend is seen when only including the X-ray detected quasars for which we have directly measured the obscured fraction from X-ray spectral fitting (magenta circles in Figure \,\ref{fig:ObsFracSFRs}); however, the obscured fraction is very uncertain at the largest SFR bin due to it only containing six sources.

\subsection{Host galaxy obscuration of highly star-forming quasars} \label{subsec:ISMNH}

The results of Section\,\ref{subsec:ObsFracSFR} suggest that the host galaxy can significantly contribute to the total quasar obscuration in highly star-forming systems. This section aims to quantify the host-galaxy ISM contribution to the total quasar obscuration using ALMA submm size constraints for the 24 submm-IR quasars. 

We calculate the column density associated with the gas in the host galaxy, taking an approach already used in previous works \citep[e.g.,][]{2019Circosta}. To do this, we first calculate the gas mass of the host galaxy, which is estimated as $M_{\rm gas} = M_{\rm H_2} + M_{\rm HI}$. The molecular gas mass is calculated from the $\rm CO(1-0)$ luminosity as $M_{\rm H2} = \alpha_{\rm CO} L'_{\rm CO}$, where $\alpha_{\rm CO}\approx 0.8 \rm \: M_{\odot}/(\rm K\: km \: s^{-1} \: pc^2)$ for starburst galaxies \citep[e.g.,][]{2012Magdis,2013CarilliWalter,2018CalistroRivera,2021Birkin}. We calculate $L'_{\rm CO}$ using the well-measured empirical relationship to the rest-frame $850\rm \: \mu m$ luminosity: $L'_{\rm CO} [\rm K\: km \: s^{-1} \: pc^2] =3.02 \times 10^{-21} \:L_{\rm 850 \mu m} [\rm erg\: s^{-1} \: Hz^{-1}]$ \citep{2016Scoville}.  We estimate the rest-frame $L_{\rm 850 \mu m}$ from the best-fitting SF component in our SED fitting, removing any AGN contribution; when the SFR is unconstrained, we can only provide an $L'_{\rm CO}$ upper-limit, which we calculate using the 99th-percentile of $L_{\rm 850 \mu m}$ distribution. We note that this relation is valid for galaxies with stellar masses $M_{\star} > \rm 2\cdot 10^{10} \: \rm M_{\odot} $, which is well within the range of our submm-IR quasars and SMGs (see Figure\,\ref{fig:prop}). We then calculate the atomic gas mass following theoretical models, which predicts $M_{\rm HI}\approx [2-5] \cdot M_{\rm H_2}$ \citep{2011Lagos,2012FuJ} for the redshifts and stellar masses of our quasars; hence, we adopt a mean value of $M_{\rm HI}=2.5 \cdot M_{\rm H_2}$. The gas masses of the IR quasars and SMGs range $M_{\rm gas, submm-IR\: quasars}=(0.04-13)\rm \: \times 10^{10} \: M_{\odot}$ and $M_{\rm gas, SMGs}=(0.3-20)\rm \: \times 10^{10} \: M_{\odot}$, respectively.

To calculate the average ISM column density, we adopt a uniform spherical geometry. Assuming that half of the gas resides inside $R_e$, the numeric density of hydrogen atoms is calculated as $n_{\rm H}= 1/2 M_{\rm gas}/(m_{\rm H} \: 4/3 \pi R_{\rm e}^3)$, where $m_{\rm H}$ is the Hydrogen mass. Then, $\langle N_{\rm H,ISM} \rangle=\int_0^{R_e} n_{\rm H} \: \rm dr = 1/2 M_{\rm gas}/(m_{\rm H} \: 4/3 \pi R_{\rm e}^2) $. All the quantities calculated in this section are reported in Table\,1 of the Supplementary Material.

\begin{figure}
\centering
\includegraphics[trim={0 0 3cm 1.7cm},clip,scale=0.45]{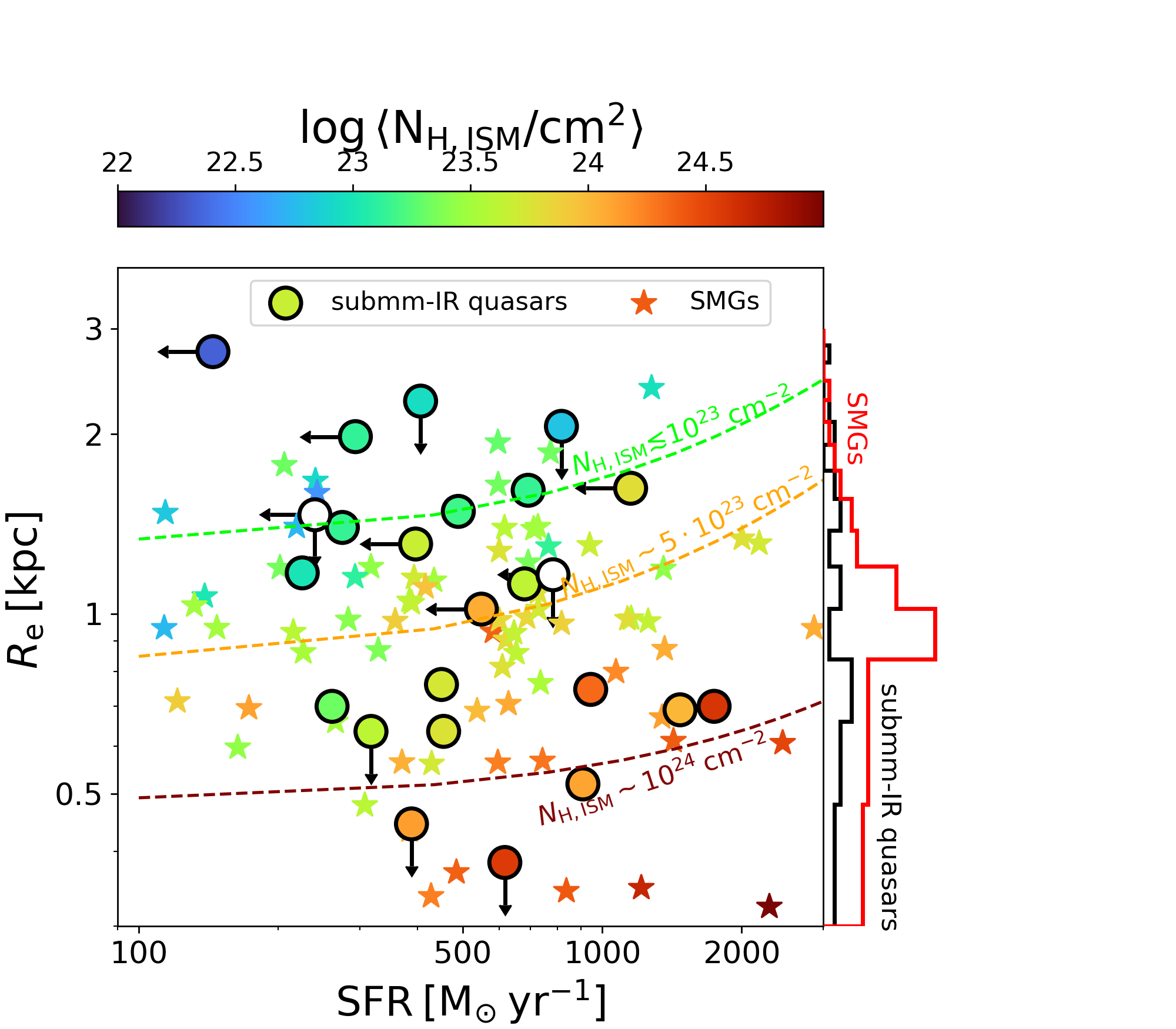}

\caption{Submm host galaxy size ($R_{\rm e}$) against the SFR, coloured by the mean ISM column density ($\langle N_{\rm H, ISM}\rangle$). Stars correspond to SMGs, and circles correspond to submm-IR quasars. SFR upper limits are plotted at 99\% confidence. Empty circles are quasars with an upper limit on $R_{\rm e}$ and SFR; hence, no ISM column density can be calculated. The dotted lines represent linear regressions between $R_{\rm e}$ and SFR for sources with $\langle N_{\rm H, ISM}\rangle < 3\cdot 10^{23} \: \rm cm^{-2}$ (green), $\langle N_{\rm H, ISM}\rangle = [3,10]\cdot 10^{23} \: \rm cm^{-2}$ (orange), and $\langle N_{\rm H, ISM}\rangle > 10^{24} \: \rm cm^{-2}$ (red). The text next to each dotted line is the mean $\langle N_{\rm H, ISM}\rangle$ in each bin. The histograms on the right side of the plot show the size distributions of the submm-IR quasars (black) and SMGs (red).} 
\label{fig:Re_SFR}
\end{figure}

Figure\,\ref{fig:Re_SFR} shows the effective radius versus SFR of our sources colour-coded by $\langle N_{\rm H,ISM} \rangle$. We find that the submm sizes of the submm-IR quasars and SMGs are in broad agreement, with values ranging $R_e \approx 0.5-2.7 \rm \: kpc$ (mean and standard error of $R_{\rm e,submm-IR\: quasars}=1.1\pm0.6\rm \: kpc$ and $R_{\rm e,SMG}=0.95\pm0.43\rm \: kpc$). Additionally, the ISM obscuration values of the submm-IR quasars range $\langle N_{\rm H,ISM} \rangle_{\rm submm-IR quasars}=(0.01- 4)\times 10^{24} \: \rm cm^{-2}$, with a mean value consistent with that of the SMGs (mean and standard error of $\langle N_{\rm H,ISM} \rangle_{\rm submm-IR quasars}=(0.8\pm 0.1)\times 10^{24} \: \rm cm^{-2}$ and $\langle N_{\rm H,ISM} \rangle_{\rm SMGs}=(1.5\pm 3.5)\times 10^{24} \: \rm cm^{-2}$), confirming that the only significant difference between the two populations is the presence or absence of quasar activity (see also Figure\,\ref{fig:prop}).

Figure\,\ref{fig:Re_SFR} also shows the results of the linear regressions between $R_{\rm e}$ and SFR for three different ISM obscuration bins, combining the submm-IR quasars and SMGs. We find that CT ISM obscurations only occur in very compact host galaxies with $R_{\rm e}\lesssim 0.8 \rm \: kpc$. For larger host galaxy sizes, even for extremely high SFRs ($>1000 \: \rm M_{\odot} \: yr^{-1} $), the ISM obscuration does not exceed $\langle N_{\rm H,ISM}\rangle \approx 5\cdot 10^{23} \rm \: cm^{-2}$. Similarly, galaxies with $R_{\rm e}\sim 2\rm \: kpc$ typically have $\langle N_{\rm H,ISM}\rangle \lesssim 10^{23} \rm \: cm^{-2}$, irrespective of the SFR value. These results indicate that the ISM can have an important contribution to the total obscuration of the system in extremely compact star-forming galaxies and that its contribution increases towards smaller host galaxy sizes.

Some studies suggest that the dust continuum sizes from submm observations can underestimate the gas sizes by a factor of $\approx 1.3-2$ \citep[e.g.,][]{2018CalistroRivera,2022Popping}, which would imply an overestimation of our ISM column densities. We find that if we increase the host galaxy sizes by a factor of two, the ISM column densities decrease by a factor of four but not changing the main results of the paper, since the majority of the submm-IR quasars have inferred $\langle N_{\rm H,ISM}\rangle > 10^{23} \rm \: cm^{-2}$. Furthermore, we note that modelling the ISM with a spherical geometry is a simplification; for example, there are eight submm-IR quasars in the spectroscopic catalogue of broad-lines (BL) quasars from \citet{2013Rosario}, indicating that we can have an unobscured view of the accretion disk due to the patchiness and clumpiness of the obscuring material. If we instead adopt a disk with a covering factor of 50\%, our ISM column densities would increase by a factor of 2. 

\begin{figure}

\centering

\includegraphics[trim={1.2cm 0 0 1cm},clip,scale=0.35]{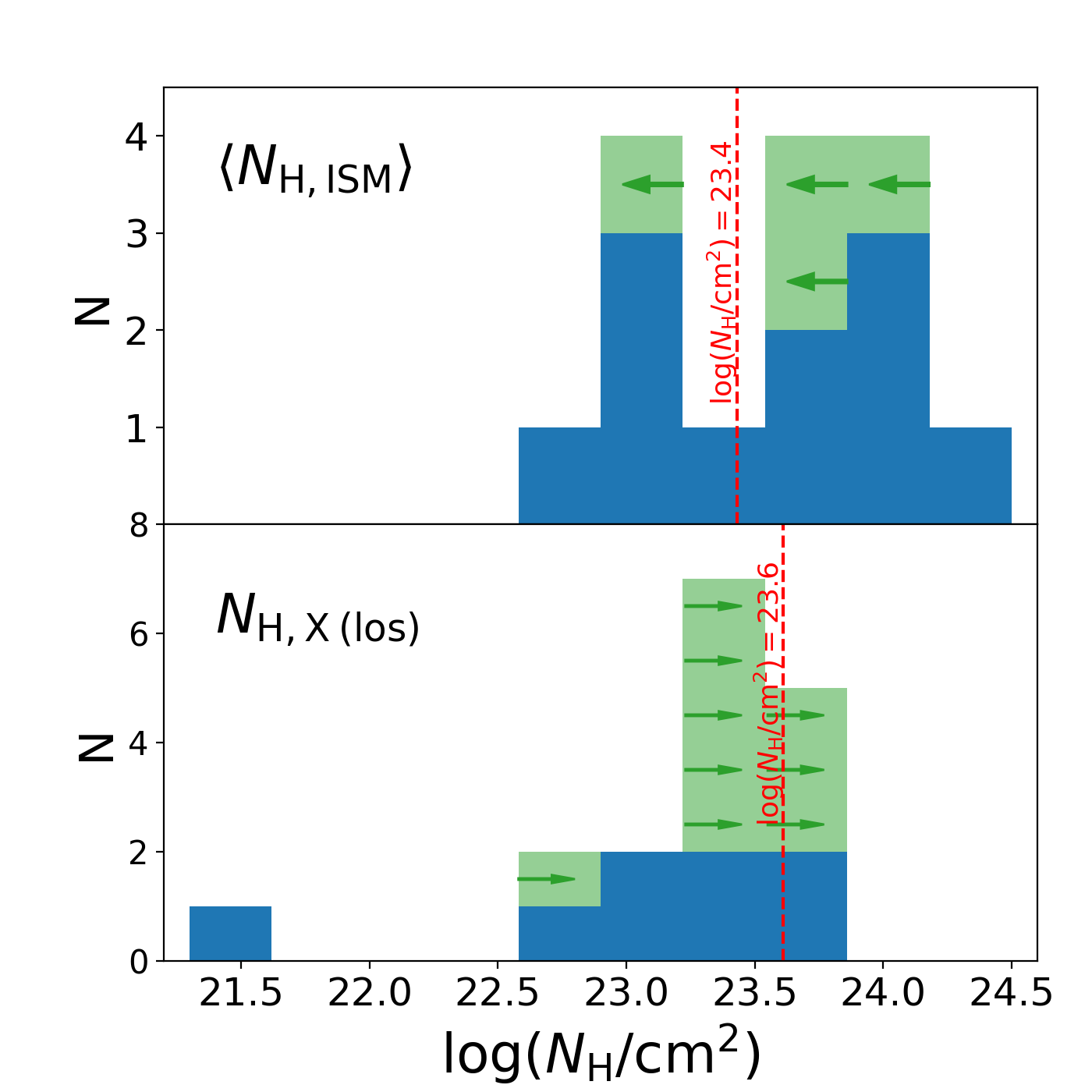}

\caption{Column-density distributions of the submm-IR quasars. {\it Top panel}: distribution of the mean ISM column density ($\langle N_{\rm H, ISM}\rangle$). The green histogram shows sources with a $\langle N_{\rm H, ISM}\rangle$ upper limit, due to an unconstrained star formation component in the SED. {\it Bottom panel}: distribution of the l.o.s X-ray column density ($N_{\rm H,X\:(los)}$). The green histogram represents $N_{\rm H,X\:(los)}$ lower limits for the X-ray undetected sources. The red dotted lines show the mean $N_{\rm H}$ of each distribution; see section \,\ref{subsec:ISMNH}.}

\label{fig:NHcomp}
\end{figure}

The X-ray spectra provide direct constraints of the l.o.s column densities ($N_{\rm H,X (los)}$) towards the central SMBH. Although we cannot directly compare these values with the estimates of the ISM column densities (which are instead average values), it is insightful to compare both distributions since they will give us some indication of the ISM contribution to the overall quasar obscuration. Figure\,\ref{fig:NHcomp} shows the distributions of $\langle N_{\rm H,ISM} \rangle$ and $N_{\rm H,X (los)}$. We exclude the BL submm-IR quasars since we know that, by definition, they have a preferential and unrepresentative unobscured l.o.s view. We also calculate the mean $\langle N_{\rm H,ISM} \rangle$ and $N_{\rm H,X (los)}$ of our sample by adopting uniform distributions when we only have limits: for the ISM column density upper limits, we set the lowest possible value to $ \log \langle N_{\rm H,ISM}/\rm cm^{2} \rangle= 21.7 $, following the results of this section; while for X-ray undetected sources, we set the highest possible value of $\log (N_{\rm H,X (los)}/\rm cm^{2})=24.7 $, following \citet{2005Alexander}. We find that the mean ISM and X-ray column density values of our sample are $\log \langle N_{\rm H,ISM}/\rm cm^{2} \rangle=23.4 \pm 0.1$ and $\log (N_{\rm H,X (los)}/\rm cm^{2})=23.6\pm 0.1$, respectively, implying that the ISM is comparable to (and potentially dominates) the overall obscuration towards the quasar in many systems. As expected, the BL submm-IR quasars have a substantially lower X-ray column density ($\log (N_{\rm H,X (los)}/\rm cm^{2})=23.1\pm 0.1$), despite their $\langle N_{\rm H,ISM} \rangle$ remaining consistent to the overall sample, demonstrating that their l.o.s column density is not representative of the typical column density towards these sources. We caution that the ALMA observations of AGN in COSMOS are biased toward X-ray sources; hence, the fraction of BL submm-IR quasars (8/24) might be overestimated.

\section{Discussion and Conclusions}

Overall, in this letter we have two key results: (1) a relationship between obscuration and SFR for IR quasars, showing that the host galaxy is likely to make a significant contribution to the increase in the obscured quasar fraction at $\rm SFR\gtrsim 300 \: M_{\odot} \: yr^{-1}$ (see Figure\,\ref{fig:ObsFracSFRs}); and (2) estimates of the ISM column densities which shows that quasars can be heavily obscured by their host galaxies (see Figure\,\ref{fig:Re_SFR}). The first result is in agreement with \citet{2015Chen}, who found that the obscured quasar fraction increases from 0.3 to 0.7 for $\rm SFR=200-630 \rm \: M_{\odot} \: yr^{-1}$. While our second result is in agreement with previous works which found that AGN host galaxies can obscure the central SMBH \citep[e.g.,][]{2017Buchner,2023Silverman} and can even be CT in some extreme cases \citep[e.g.,][]{2019Circosta, Gilli22}. However, our study combines these two results for the first time by exploring the connection between the host galaxy obscuration and the SFR in IR quasars. In this final section, we discuss our results and their implications for quasar models.

\subsection{ISM contribution to the obscured quasar fraction}

Section \,\ref{subsec:ISMNH} shows that submm-IR quasars and SMGs can be heavily obscured by the ISM and even CT for the most compact sources. Such high levels of host galaxy obscuration are not unexpected. Indeed, SMGs are known to have very high gas surface densities, with typical values of $\rm \Sigma_{gas}\approx (10^2-10^4) \: M_{\odot} \: pc^{-2}$ (\citealp[e.g.,][]{2010Daddi,2013Tacconi,2021Kennicutt}, equivalent to $\rm N_{\rm H} \approx 10^{22}-10^{24}\: cm^{-2}$), and high ISM column densities, with typical values of $N_{\rm H}=9.8_{-0.7}^{+1.4}\times 10^{23} \rm \: cm^{-2}$ \citep[e.g.,][]{2017Simpson}. Extreme host galaxy obscuration has also been found in the local ultraluminous IR galaxies NGC 4418 and Arp 220, which are estimated to be obscured by $N_{\rm H}\sim 10^{25}-10^{26} \rm \: cm^{-2}$ \citep[e.g.,][]{2015Barcos,2020Dwek,2021Sakamoto}.

Large ISM column densities are also predicted in simulations. For example, \citet{2019Trebitsch} found that during periods of rapid growth, the gas at galactic scales has a column density at least comparable to the gas at nuclear scales. Similarly, \citet{2011Bournaud} finds that disk instabilities can trigger mass inflow towards the central engine and that during this phase, the total column density is dominated by the gas over galactic scales, which can achieve $\rm \langle N_{\rm H, ISM}\rangle > 10^{23}\: cm^{-2}$, and even reach CT levels. However, in these simulations, unobscured l.o.s. are still expected towards the central engine 10\% of the time, in agreement with the broad-line (unobscured) quasars residing in starburst galaxies found in our work.

We note that the submm sizes of the submm-IR quasars hosts and SMGs are much more compact than those of massive main-sequence galaxies at similar redshifts, ranging between $\approx 3-6 \rm \: kpc$ \citep[e.g.,][]{2020Kaasinen}, which implies typical ISM column densities of $\langle N_{\rm H,ISM} \rangle\approx (1-9)\cdot 10^{21} \: \rm cm^{-2}$ for comparably massive main sequence galaxies with $\rm SFR\approx 50-90 \rm \: M_{\odot} \: yr^{-1}$. Assuming that the quasars living in normal star-forming galaxies have submm sizes comparable to typical main-sequence galaxies, we would therefore expect their ISM column densities to be typically unobscured and, therefore, to have a negligible contribution to the quasar obscuration, particularly for $\rm SFR\lesssim 100 \: M_{\odot} \: yr^{-1}$, consistent with our results.

\begin{figure}

\centering

\includegraphics[trim={0cm 1cm 1cm 0.8cm},clip,scale=0.18]{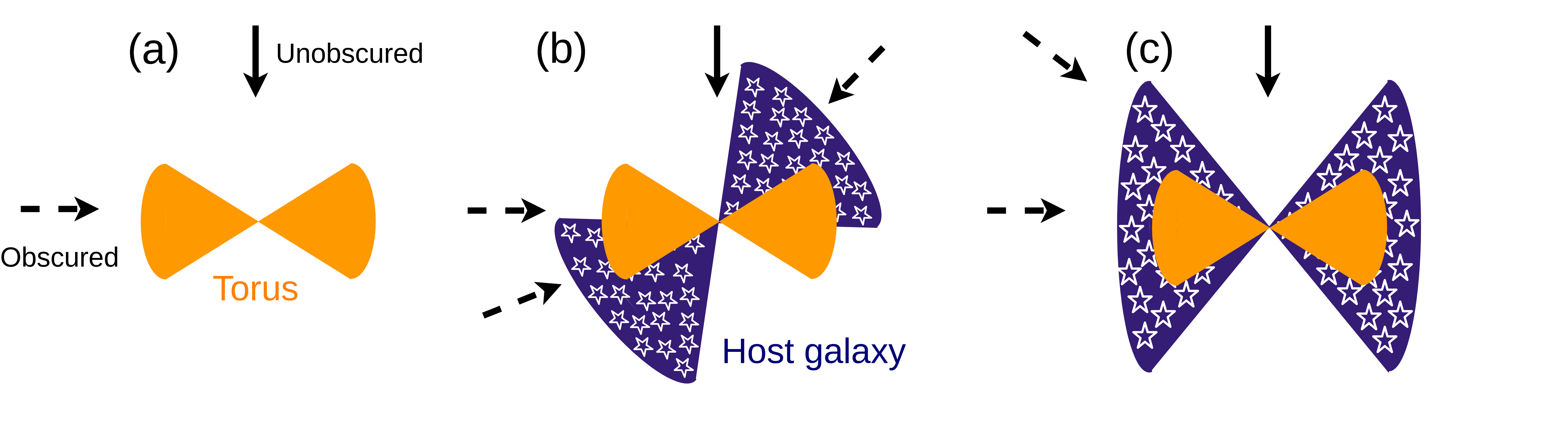}

\caption{Illustration of the dominant sources of obscuration. Panels (a), (b), and (c) show different relative orientations between the torus (orange flared disk) and host galaxy (blue flared disk with stars). The dashed and continuous arrows represent obscured and unobscured l.o.s, respectively. The cartoons are illustrative, and the sizes and opening angles are not drawn to scale. }
\label{fig:obs_sketch}
\end{figure}

Overall, our results indicate that a large component of the quasar obscuration can come from the ISM, in addition to the obscuration from the dusty torus. Figure\,\ref{fig:obs_sketch} illustrates three scenarios of quasar obscuration along different lines of sight. At $\rm SFR\lesssim 300 \: M_{\odot} \: yr^{-1}$, the obscured quasar fraction remains constant around $\approx 0.65$; hence, it is reasonable to assume that the torus dominates the obscuration for these systems (panel a). Assuming no increase in the torus covering factor with SFR, the increase in the obscured quasar fraction to $\approx 0.8-0.9$ would be due to the ISM, implying that the ISM solely obscures $\approx 10-30\%$ of the quasars with $\rm SFR\gtrsim 300 \: M_{\odot} \: yr^{-1}$ (see panels b and c). Taking a basic probabilistic approach and assuming that the orientation of the torus and the ISM are independent (see panel b), the ISM dust-covering fraction would be $\approx 0.4-0.7$. However, if there is some co-alignment (see panel c), then the ISM dust-covering fraction could be as high as $0.9$ in the highest SFR systems.

\subsection{Implications of our results for quasar models}

The similarity in the host galaxy properties of the SMGs and submm-IR quasars, together with their large estimated ISM column densities, suggest that a simple unification model postulating that the obscuration is only due to the dusty torus does not hold for these extreme systems. It has been postulated that SMGs (and consequently, submm-IR quasars) may be experiencing a compaction phase where the ISM material is falling towards the centre of the galaxy, triggering a compact starburst, and subsequently activating the SMBH, which is supported by the fact that both populations have similarly compact submm sizes. This phenomenon, also known as {\it 'wet compaction'}, is triggered by an extreme loss of angular momentum caused by either violent disc instabilities, counter-rotating streams or wet mergers \citep[e.g.,][]{2014Dekel, 2015Zolotov, 2023Lapiner}. A number of studies have proposed {\it 'wet compaction'} as the explanation for the high fraction of compact star-forming galaxies hosting an AGN, as compared to extended co-eval star-forming galaxies with similar stellar masses and redshifts \citep[e.g.,][]{2017Kocevski, 2020Chang, 2019Silverman, 2021Lamperti}, an explanation also supported by cosmological simulations \citep[e.g.,][]{2019Habouzit,2023Lapiner}. 

Recent results have shown that AGN in mergers are typically heavily obscured  \citep[e.g.,][]{2017Ricci}, that obscured AGN/quasars reside in preferential close galaxy pairs and massive large scale environments \citep[e.g.,][]{2023Dougherty,2023Petter}, and have also excess radio emission over that found for unobscured quasars (\citetalias{2022Andonie}), consistent with that expected from quasar-driven outflows and jets interacting with the host-galaxy ISM (i.e.,\ ``AGN feedback'', \citealp[e.g.,][]{2019Klindt,2020Fawcett,2021CalistroG}). Our study extends these differences to the star-forming environments, showing that obscured quasars are preferentially found in compact starbursts. Combined, these suite of results can be used to argue that different quasar sub populations represent different phases in the overall evolution of quasars \citep[e.g.,][]{2014Lapi,2016Mancuso,2018Lapi}. A key uncertainty is the submm sizes of the lower SFR quasars ($\rm SFR\lesssim 100 \: M_{\odot} \: yr^{-1}$), which can be used to determine their relationship with the high SFR quasars studied here (e.g.,\ comparably extended to co-eval main-sequence galaxies versus compact like the high SFR quasars). A combined ALMA and radio analysis of IR quasars across the full range of SFR would provide key insights into the relationship between different quasars sub populations and test quasar evolutionary models.

\section*{Acknowledgements}

This work has been supported by the EU H2020-MSCA-ITN-2019 Project 860744 “BiD4BESt: Big Data applications for black hole Evolution STudies”. DMA and AP thank the Science Technology Facilities Council (STFC) for support from the Durham consolidated grant (ST/T000244/1). AP acknowledges partial support by STFC through grant ST/P000541/1. CH acknowledges funding from an United Kingdom Research and Innovation grant (code: MR/V022830/1). AL is partly supported by the PRIN MIUR 2017 prot. 20173ML3WW 002 ‘Opening the ALMA window on the cosmic evolution of gas, stars, and massive black holes’. DJR acknowledges the support of STFC grant NU-012097

\section*{Data Availability}

The datasets generated and/or analysed in this study are available from the corresponding author on reasonable request.



\bibliographystyle{mnras}
\bibliography{references}







\bsp	
\label{lastpage}
\end{document}